\documentclass[]{interact}

\usepackage{epstopdf}
\usepackage[caption=false]{subfig}

\usepackage[numbers,sort&compress]{natbib}
\bibpunct[, ]{[}{]}{,}{n}{,}{,}
\renewcommand\bibfont{\fontsize{10}{12}\selectfont}
\makeatletter
\def\NAT@def@citea{\def\@citea{\NAT@separator}}
\makeatother

\theoremstyle{plain}

\theoremstyle{definition}

\theoremstyle{remark}

\begin{document}

\title{First Light: Switching on Stars at the Dawn of Time}

\author{
\name{E. Chapman\textsuperscript{a}\thanks{Chapman E. Author. Email: emma.chapman@nottingham.ac.uk}}
\affil{\textsuperscript{a}The Centre for Astronomy and Particle Theory, The School of Physics and Astronomy, University of Nottingham, University Park, Nottingham, NG7 2RD, UK}
}

\maketitle

\begin{abstract}
The Era of the First Stars is one of the last unknown frontiers for exploration: a poorly understood billion years missing from our cosmological timeline. We have now developed several methods for finally filling in the lost billion years of the history of our Universe: stellar archaeology, detecting primordial hydrogen using 21 cm cosmological emission, and observing the earliest galaxies, most recently using the James Webb Space Telescope. This review will summarise why the first stars and galaxies are unique and worthy of observation, and the methods employed by the groundbreaking telescopes aiming to detect them.

\end{abstract}

\begin{abbreviations}
LOFAR: Low Frequency Array | SKA: Square Kilometre Array | MWA: Murchison Widefield Array | EDGES: Experiment to Detect the Global EoR Signature | EoR: Epoch of Reionisation | JWST: James Webb Space Telescope

\end{abbreviations}


\section{Introduction}

Our understanding of stars, both in composition and ancestry, has developed considerably over the ages of humanity. Early cave paintings depict constellations as animals, while stars were the essence of Babylonian deities, or the lamps hung by Egyptian Gods. The Ancient Greeks more accurately saw them as celestial bodies, made of the last of their four elements of the cosmos: earth, water, air and fire \cite{pannekoek89,dreyer53}. Most of these ancient cosmologies placed the stars close by Earth, though in a different sphere or spiritual plane. When Galileo peered through his telescope, though, he found that stars still looked like point sources, implying gargantuan distances from Earth.

Clearly, these bodies were a power source greater than any we knew on Earth. In the late nineteenth century, there were hypotheses of meteors feeding an insatiable globe of fire, and heat produced through gravitational contraction. While gravitational contraction does release energy, and a star is undoubtedly under enormous gravitational pressure, this process would take only ten or so million years to complete – not long enough for a planet to form, let alone a human. Through the early twentieth century, scientists considered chemical explanations, but as George Gamow put it in 1949, “If the Sun were made of pure coal and had been set afire at the time of the first Pharaohs of Egypt, it would by now have completely burned to ashes” \cite{gamow40}. After the discovery of radioactivity in uranium in 1896, and then radium\footnote{Radium was the wonder metal of the age. Toted as a cure for most medical ailments, the market was flooded with radium food, beauty products and even contraceptives. Of course, dosing yourself with radioactivity daily would never have a good outcome, and soon enough, we found instead the wonder metal to be our kryptonite. See ‘The Radium Girls’\cite{moore16} for an astonishing account of the fight to recognise the fatal damage radium could do.}, physicist Ernest Rutherford and radiochemist Frederick Soddy began investigating thorium. Noting that it naturally generated enough heat to melt its own weight in ice every hour, they suggested the Sun could be powered by similar radioactive processes \cite{soddy09}, though they were stumped how this could be so. The Sun appeared to be comprised of the same elements as the Earth, and the levels of uranium were nowhere enough high enough to account for the heat and radiation emitted by the Sun. Astronomer Arthur Eddington suggested that the internal source of energy was indeed atomic, but came from the “transmutation of elements”\cite{eddington26}. 

It was not clear what could be transmuting, though. Until 1925, the main consensus was that the Sun had the same chemical composition as Earth. Scientists could determine the content of stars using a method called \textit{spectroscopy}. In one of the more surreal moments of my life, I once explained the method to the captivated (not captive, I promise) musicians making up the UK Pink Floyd Experience. The famous logo on their t-shirts shows a glass prism splitting white light into the colours of the rainbow: a spectrum. Isaac Newton performed a crucial experiment in 1666 that proved that white light was made up of different colours \cite{newton1704}. While others had previously produced spectra using prisms, it was thought that the pure light was contaminated by the air. Newton was able to isolate single colours, and show that passing, for example, blue light through a second prism did not produce a rainbow... it produced blue light. Light is made up of an array of different wavelengths. By the turn of the nineteenth century, optics experts such as William Hyde Wollaston and Joseph von Fraunhofer had produced impressively pure lenses that split light into fine wavelengths. Mysterious gaps, that had been below the resolution limit, popped out. Exposing different elements to light in laboratories produces identical and individual patterns of dark gaps, akin to barcodes. Atoms of a particular element will absorb
light at wavelengths relating to the energy levels that electrons can occupy within that atom. The gaps between these energy levels differ between elements, and therefore so do the resulting absorption lines. Decoding and separating the forest of barcodes within the solar spectra reveals, amongst others, hydrogen, helium, iron, calcium, sodium, and magnesium. These elements identify their presence by multiple absorption lines, according to their ionization state, which depends on properties such as density, temperature and the pressure we subject the element to. 

Cecilia Payne-Gaposchkin’s 1925 thesis \cite{payne25} was the first to work through the mathematics of ionization theory when applied to stellar spectra. She found that the strength of a spectral line was not directly related to the abundance of the element causing it. Instead, it indicated the ionization state of the element. When Payne-Gaposchkin explored this further, she found that the abundances of elements heavier than hydrogen were astonishingly lower than previously thought, to the point where hydrogen was one million times more abundant than the elements heavier than helium. This is very different to the abundances of those elements on Earth and so, after a period of disbelief both for herself and the community, the stars were identified as celestial bodies unlike planets \cite{russell29}. To clarify some jargon, we will henceforth use 'metals' to refer to any element heavier than helium. It is just the way of astronomy - those lighter elements make up almost the entire Universe so we might as well simplify the periodic table! (Figure 1)

\begin{figure}
\centering
\includegraphics[width=0.6\linewidth]{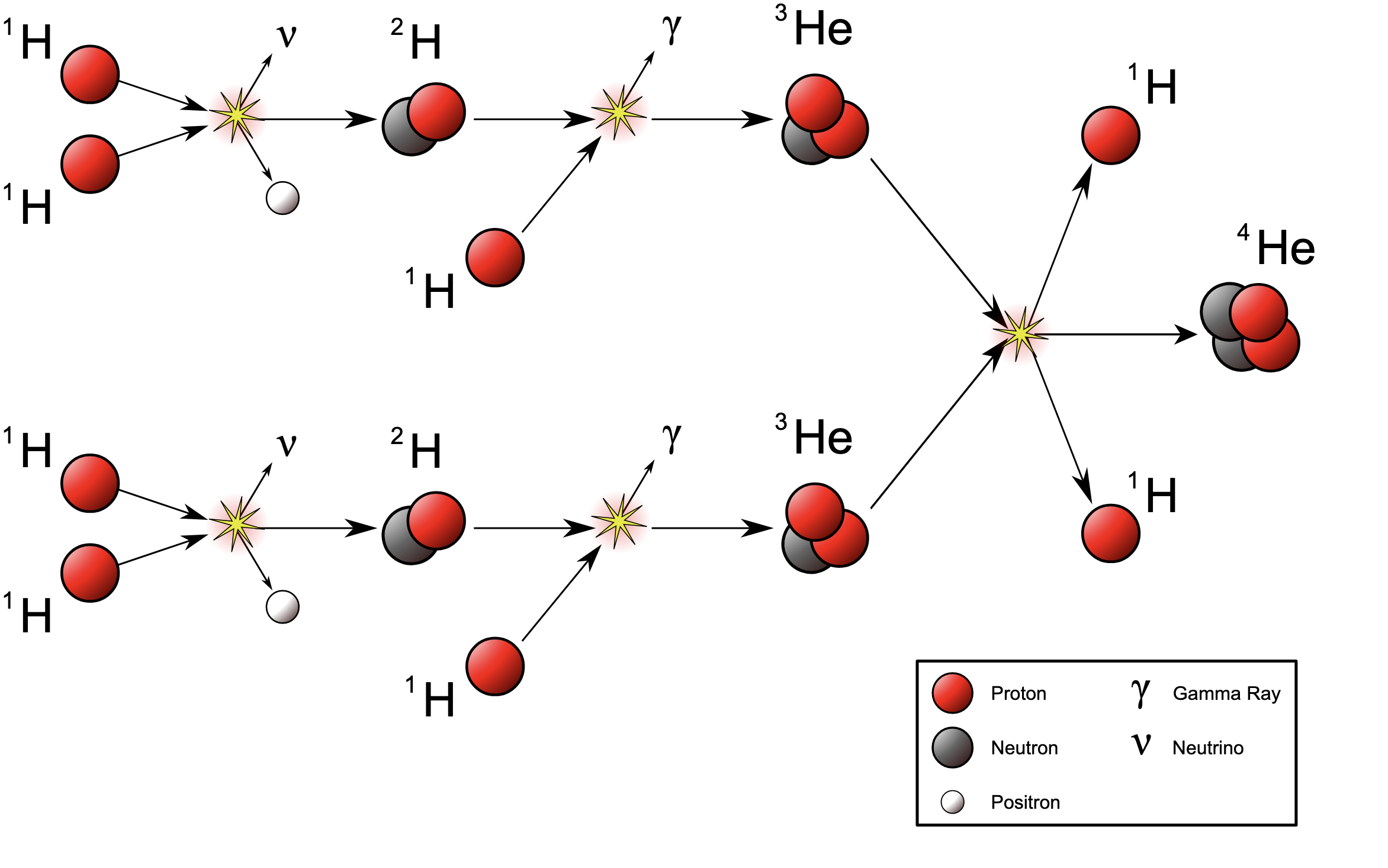}
\caption{The proton-proton chain reaction whereby hydrogen fuses into helium. This pathway is common in stars smaller than the Sun and, because of the lack of metal catalysts, the first stars.} \label{fig-fusion}
\end{figure}

The huge abundance of hydrogen left no question what was transmuting. The fusion of two hydrogen atoms into a helium atom releases energetic photons, alongside other byproducts. The energy released by a single fusion reaction is tiny, about $1\times10^{-12}$ Joules. There is a heck of a lot of hydrogen in the Sun, however, and, in its core, the Sun fuses 620 million tonnes of hydrogen every second. Finally, we have a mechanism that explains how the Sun could live for billions of years.

While the scientific community had accepted that the Sun was not the same as the Earth in terms of composition, they now stuck to the position that all stars have the same abundances as the Sun. In the later nineteenth and early twentieth century, the “Harvard computers” (a group of women including Williamina Fleming and Annie Jump Cannon, well worth reading about as a separate topic \cite{sobel_glass16}) had classified thousands of stellar spectra into different spectral classes, represented by the letters 'OBAFGKM'. But, while they noted stellar spectra varied between stars, they put the difference in the spectral line strengths down to differences in the star's temperature, not its composition. Another two decades would pass before new data challenged this assumption. In 1951, astrophysicists Joseph Chamberlain and Lawrence Aller capitalised on improvements in spectroscopy to find a star with such faint spectral lines that it indicated that the iron abundance\footnote{'abundance' here refers to the mass fraction of that element compared to hydrogen} was 1/10th that of the Sun. Once again, disbelief reigned: “the observed [iron] abundance appears to be smaller than in the sun, although this conclusion must be taken with caution“ \cite{chamber51}. The astronomical community had to come to terms with breaking another assumption: stars were not all the same.

They then divided the stars into two populations. Population I, containing solar-metallicity stars, and Population II, containing these new `metal-poor’ stars. Over the decades, we have observed stars with fainter and fainter metal spectral lines. So much so, that the sub-categories of Population II in terms of metallicity have got quite ridiculous, extending all the way through ‘very metal-poor: 1/100th solar iron abundance’, past ‘Mega metal-poor: 1/1,000,000th solar iron abundance’ to, wait for it, ‘Ridiculously metal-poor: 1/10,000,000,000th solar iron abundance’ \cite{frebel18}.

The division of these stars was not entirely arbitrary, or based on metallicity alone. Nancy Grace Roman’s work in 1950 had already hinted that metal-poor stars were preferentially found in the halo of our Galaxy\cite{roman50}. We know the halo is the retirement home to which old stars drift and so, in hindsight, the conclusion is simple: the younger the stars, the higher their metal content. Population II is the older generation, made of older gas, less polluted with the metals they would eventually themselves expel upon their deaths. As cosmic time went on, and stars died and produced new generations, the metallicity of the Universe increased.

\section{A Different Species of Star}
\subsection{Was there a `first' star?} 

The assumption that there was a first star at all depends on whether the Universe had a beginning, or has been as it is for all time. 13.8 billion years ago, the Universe came into being with a Big Bang: a brutally fast inflation followed by a steady expansion. In the late 1920s, after Lemaitre and Hubble noted that the vast majority of galaxies appeared to be moving away from us \cite{hubble29}\cite{lem31}, it was mooted that the Universe could be expanding. Within the framework of Big Bang theory came a successful sub-theory for the observed abundances of light elements \cite{alpher48}, which we have measured the proportions of to exquisite accuracy \cite{pitrou18}. And finally, the nail in the coffin for any non-believers, we successfully predicted and observed the Cosmic Microwave Background (CMB) radiation  \cite{penzias65}\cite{dicke65}\cite{planck20}. The Big Bang implies the universe had a beginning. The theory also implies that the energies and temperatures involved in the immediate aftermath were colossal, far too high for fully formed stars to pop out. They had to be built, and indeed even build the elements beyond helium. The aftermath of the Big Bang was no place for heavy elements. I like to imagine a living room full of Lego and sugar-crazed toddlers - you can try to build that tall tower but it won’t be long until one of your tiny friends collides with it, smashing it to bits. In the early Universe, everything was hot and full of energy. The Universe was too hot for fusion to be a viable method of creating heavy elements... and even if a rare reaction occurred, the products couldn’t survive for very long. 

By the mid-twentieth century, the narrative of the Universe up to the present day was mostly in place… well, the beginning and end of the story was, anyway. The Universe started with a Big Bang, producing the simplest of elements: hydrogen and helium. A good start. And they had fleshed the present day out with new characters: two populations of stars showing an evolution of metallicity. Clearly these metals were produced over time, polluting the primordial gas such that new generations of stars formed with every increasing metallicity. A sensible ending. But what about the middle of the story? (Figure 2) How did the Universe jump from pristine to polluted? We can extrapolate the missing piece with ease. Population I: metal-rich. Population II: metal-poor. Enter Population III: metal-free. There had to be a population of stars that formed directly from that primordial gas, stars that were powered by fusion just like any other stars, producing the first heavy elements, and upon their deaths releasing those elements into the surrounding gas. Astronomers searched the light of the stars over decades, trying to find stars without signs of heavy elements in their spectra \cite{bond70}. They couldn’t find any \cite{bond81}. They dug deeper and deeper, constantly pushing the limits of the technology. For example, the spectra of SMSS
J1605-1443 indicates an iron abundance 1,000,000th that of the Sun \cite{nordlander2019}, while there was no iron line observed at all in SM0313-6708, and they could place only an upper limit \cite{keller2014}. While we can explain the low iron abundance of some stars by billions of years of accretion producing a kind of camouflage, the presence and strength of other metal absorption lines rules out these stars as true metal-free stars. We are approximately 70 years on from that first discovery of a metal-poor star and, nevertheless, and we have not found a Population III star. One of the leading actors is missing, and the show makes little sense without them. Population I bathes us in youthful sunlight, the elderly Population II stars potter around us... but where is Population III?

\begin{figure}
\centering
\includegraphics[width=0.7\linewidth]{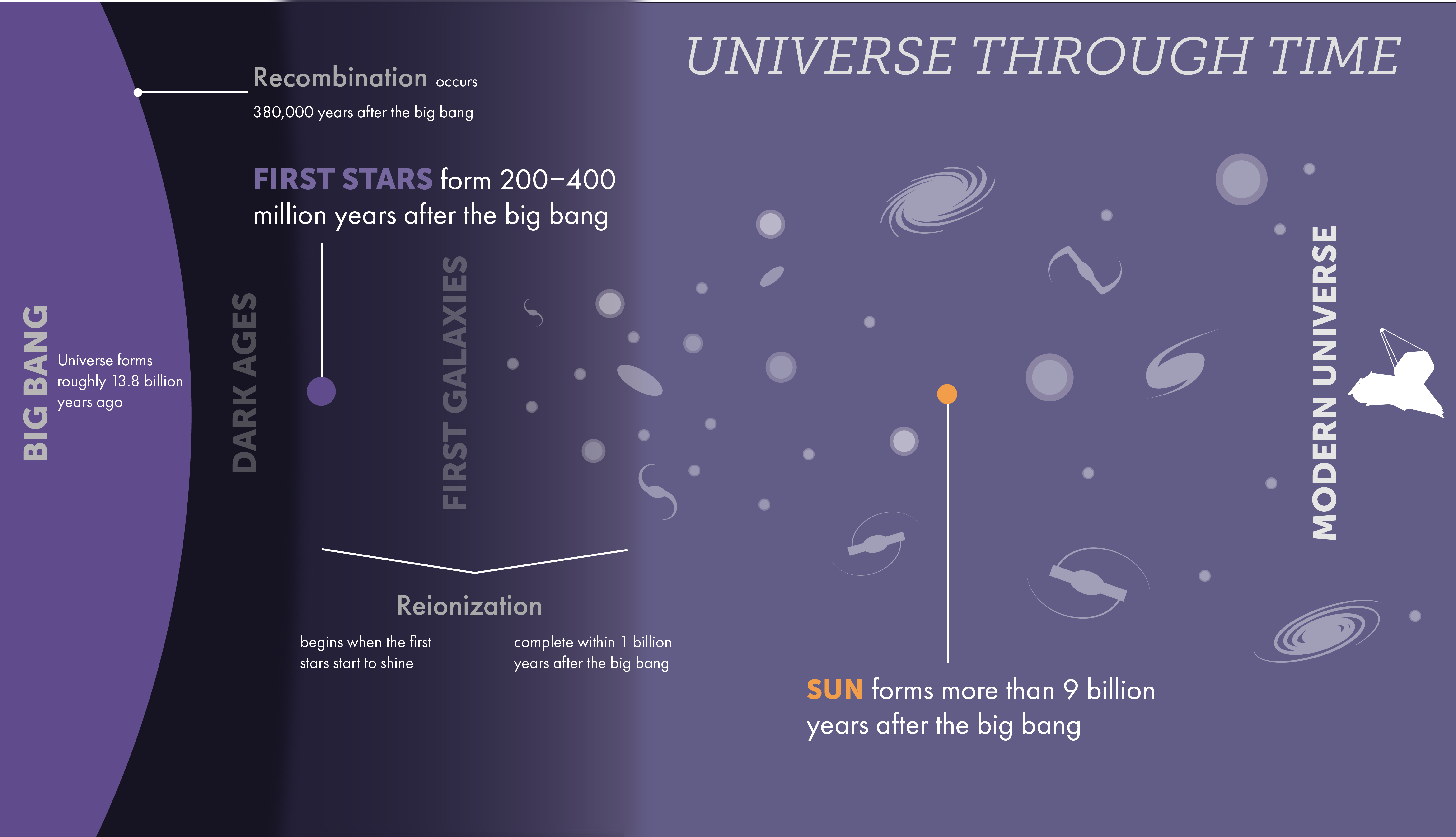}
\caption{The History of the Universe, progressing from left to right. The Universe begins with the Big Bang, an event of such energy that subatomic particles could not bond to form atoms until recombination, 380,000 years later. Then, electrons and protons can form neutral hydrogen, which coalesces through the dark ages until the highest density regions collapse to form the first stars. The neutral hydrogen is then ionized once more, during reionization, by the photons produced by the fist stars and galaxies. Image Credit:STScl} \label{fig-timeline}
\end{figure}

\subsection{A Star is Born}

Let’s dig deeper into the idea that the early Universe was ‘simple’. It would probably have seemed boring to the average optical observer, though they would observe a kind of Big Bang sunset. Hydrogen and helium are invisible in the optical at the temperatures natural to Earth. In much more energetic environments, like that of the Big Bang, those atoms would have emitted photons at very high frequencies: higher than the optical range. As the volume of the Universe increased, the temperature of the gas dropped, and the energy of the photons emitted slid down the frequency scale. At some point the emission frequencies would enter the optical, glowing white-hot, before dimming through the yellows, oranges and finally reds… setting into the radio and microwave frequencies\cite{anderson18}. Once again, just a few million years after the Big Bang, the Universe was dark, and the Dark Ages began (albeit with the Universe still at high enough temperature to incinerate our poor observer).

\begin{figure}
\centering
\includegraphics[width=1\linewidth]{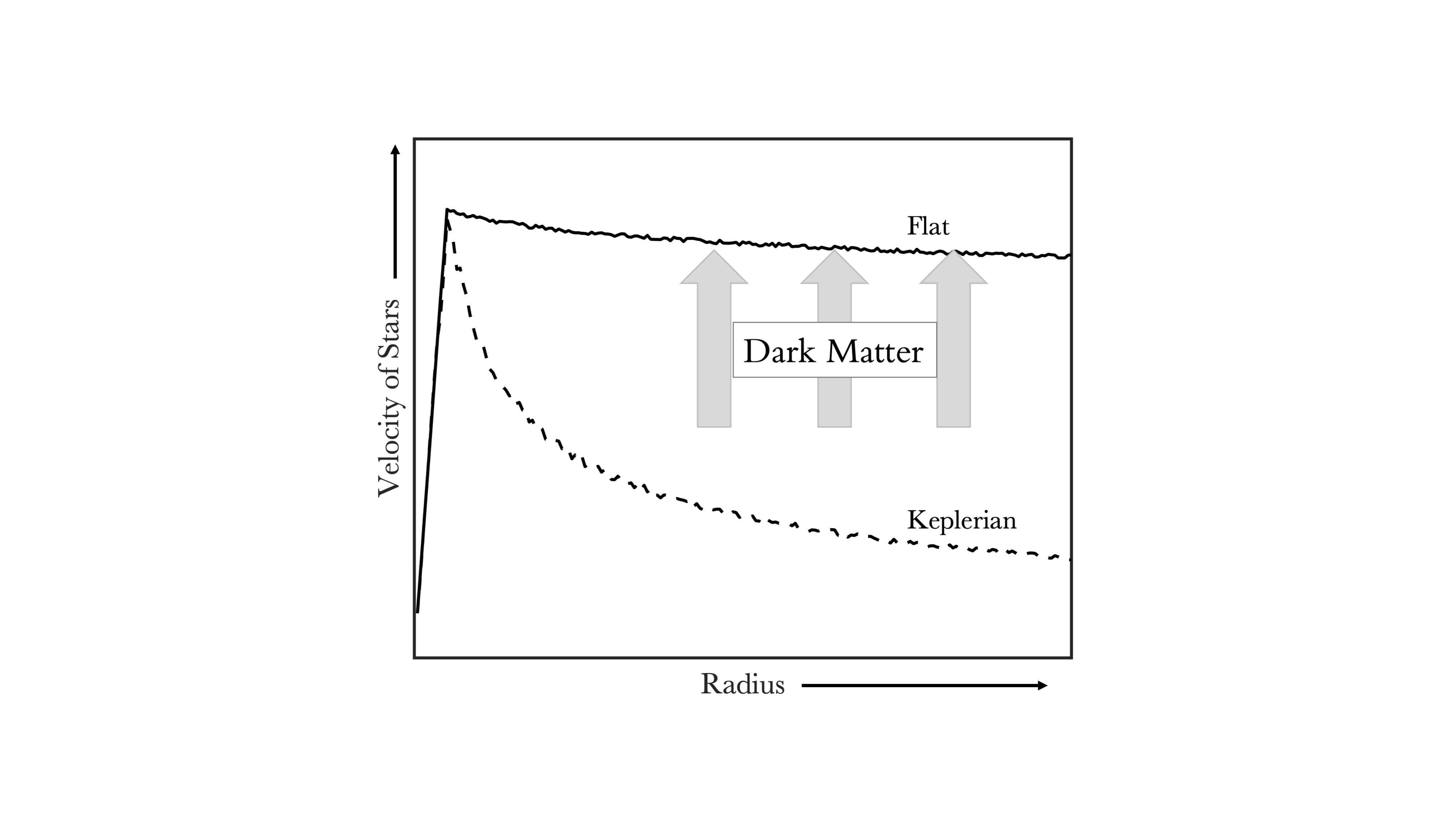}
\caption{The velocity of stars in increasingly large orbits around a galactic centre should decrease. Instead, most spiral galaxies have flat rotation curves, indicating hidden mass that does not interact with light.} \label{fig-rotcurve}
\end{figure}

Despite the calm appearance of an intermission, all is not quiet backstage. Under the surface, invisible to the observer, is a web of dark matter stretching the breadth of the Universe. The consensus is that dark matter is an exotic particle that does not interact with light, making it all-but-impossible to detect directly. We have only indirectly detected it, by noticing its gravitational effects. In the 1970s, observations showed that stars in long orbits around the centres of galaxies were moving faster than expected, in fact, just as fast as stars in much tighter orbits. We expect the velocities of orbiting stars to drop in a ‘Keplerian’ fashion, as $v(r) \propto 1/r^{\frac{1}{2}}$. Instead, stellar observations of galaxy after galaxy showed flat rotation curves \cite{bosma78, rubin82, rubin85, bhatt13} (see Fig. \ref{fig-rotcurve}). The only way\footnote{Strictly speaking, this is the ‘only way’ only if we assume Newton’s law of gravity hold at all scales. There is plenty of research into modified gravity (e.g. \cite{ishak18,nojiri17,clifton12,milgrom83}), where the laws of gravity act differently on the large scales where we observe the stars in the outer edges of a galaxy. For our purposes though, we assume Newton’s laws hold at all scales.} we can account for a flat rotation curve is to let the mass enclosed within the orbit to scale with $r$, but this is not what we observe if we determine mass based on luminosity. The luminosity (and thus stellar mass) of a galaxy drops off quickly, so to ensure $M(r) \propto r$, there must be some other mass in the outer reaches of galaxies, and it is clearly not emitting light. Thus, `dark matter'.

Without dark matter in our cosmological model, galaxy cluster members move too fast, the cosmic microwave background looks different, the whole large-scale structure of the Universe doesn’t match what we observe. The evidence points to dark matter comprising approximately 85$\%$ of the total matter in our Universe\footnote{When we consider the more general mass-energy budget of the Universe, dark matter comprises only $27\%$; ‘normal’ matter, $5\%$; and dark energy, $68\%$. Dark energy is a hypothetical form of energy that is causing the acceleration of the expansion of the Universe. Dark energy only acts on the larger scales and does not affect the formation of the first stars, as the matter density of the Universe dominates. It is only just now in cosmological time that we think dark energy has become the most important of the three mass-energy constituents.} \cite{planckcosmo}. Dark matter pervades the Universe, making itself known on the large scales of galaxies and above. Because we don’t know exactly what it is, we cannot be sure of when it formed, but the peaks and troughs of that dark matter web are already imprinted on the CMB at 380,000 years after the Big Bang. Tiny instabilities in the density field of dark matter had collapsed in on themselves, creating a web of dense filaments, separated by empty voids, and connected by dense halos of dark matter at the junctions. These halos collapse until the internal pressure of the halo balances the gravitational force pressing down. They do not carry on collapsing to form, say, a dark matter star, because a system of dark matter cannot lose energy in the same way as normal matter can. Unable to interact electromagnetically, the dark matter cannot lose energy by emitting photons, and so the internal pressure remains high, stabilising the collapse earlier than in normal matter. In this way, a scaffold is formed, and a scaffold with a strong gravitational pull on all the surrounding normal matter. The hydrogen and helium gas collapses around the filaments and halos, forming an identical cosmic web. The dense regions of gas become stellar nurseries, stimulating and nurturing star formation (Figure 4).

\begin{figure}
\centering
\resizebox*{7cm}{!}{\includegraphics[width=1\linewidth]{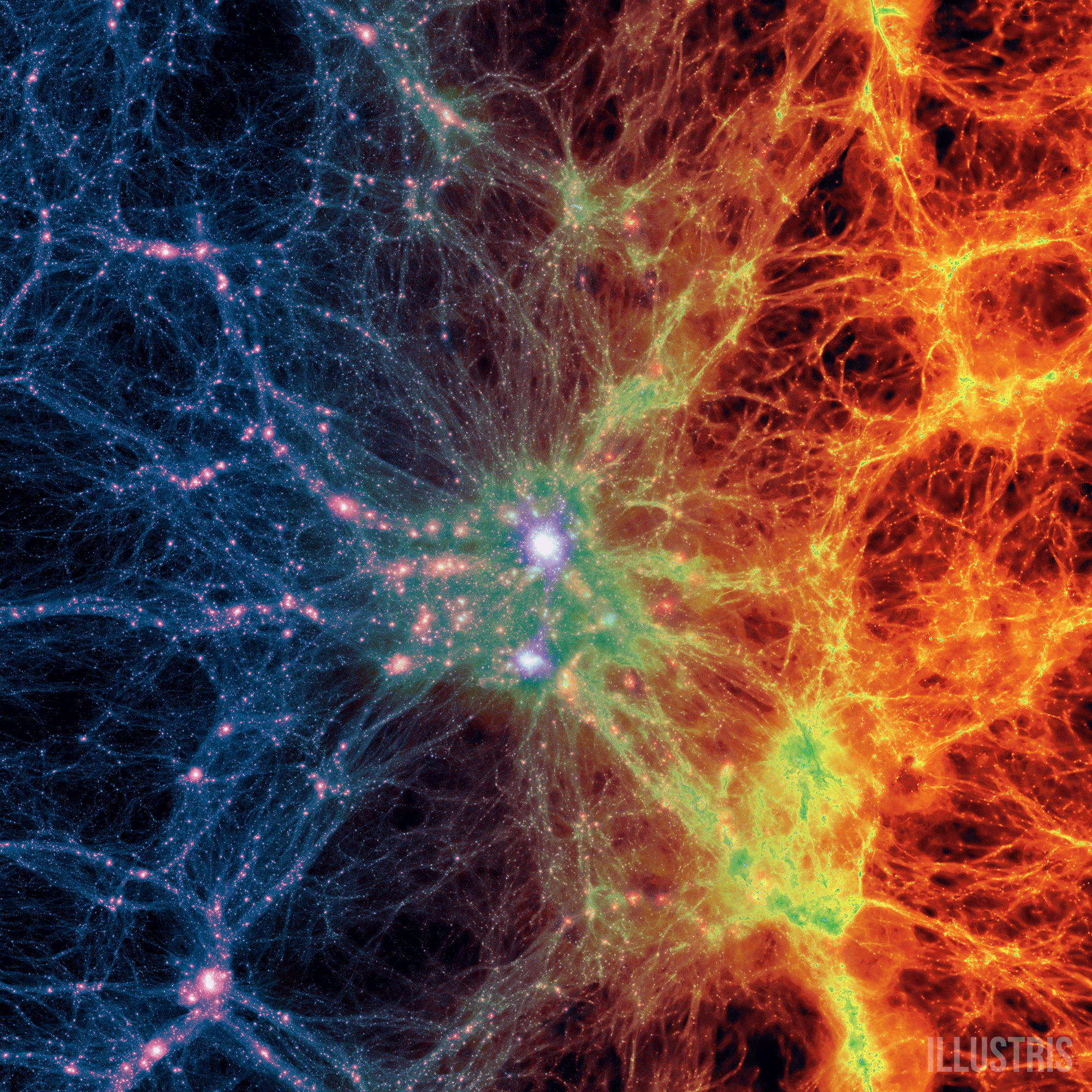}}\hspace{5pt}
\resizebox*{7cm}{!}{\includegraphics[width=1\linewidth]{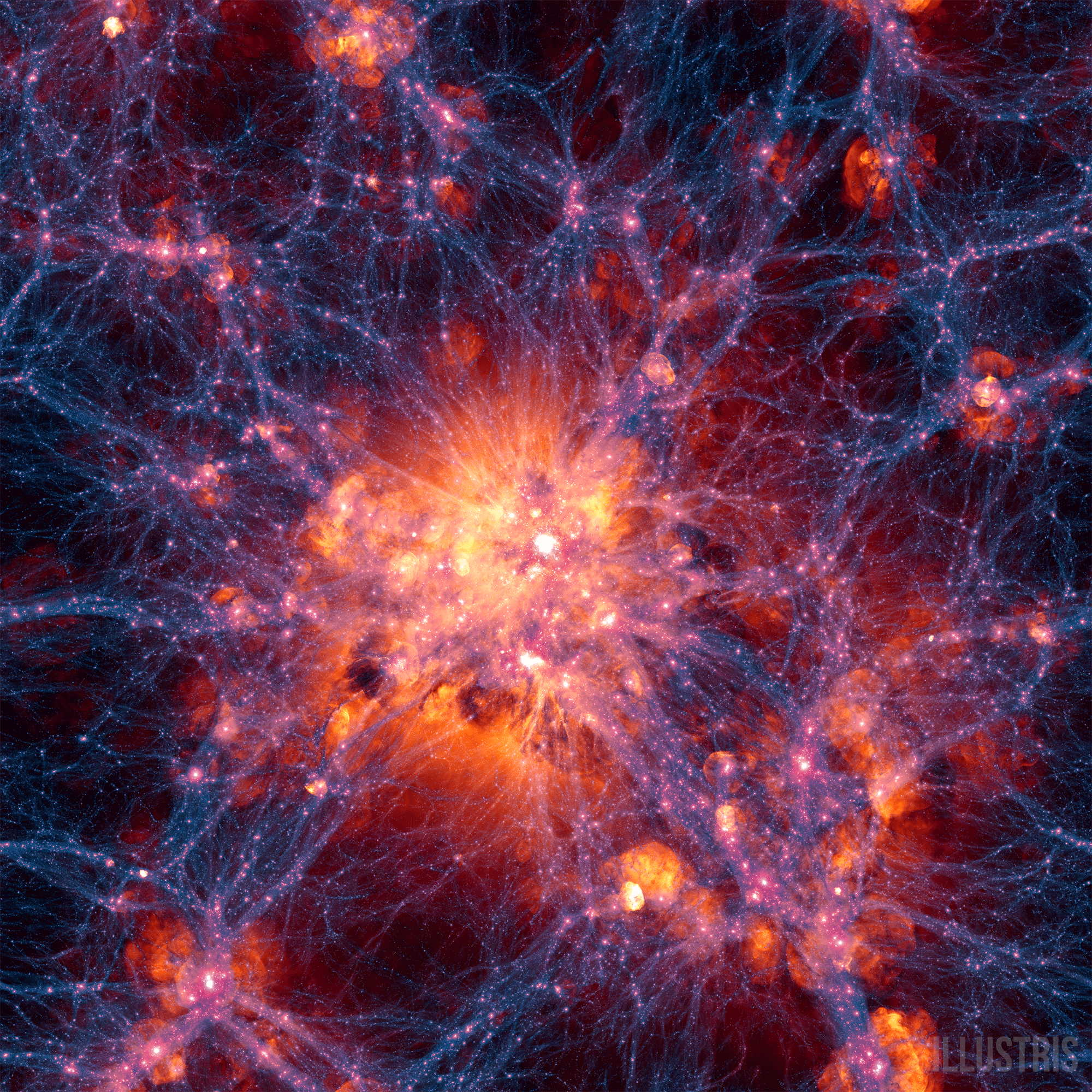}}\hspace{5pt}
\caption{A large scale projection through the Illustris volume at $z=0$, centered on the most massive cluster in the simulation, 15 Mpc/h deep. On the left we see dark matter density transitioning to gas density. On the right we see the dark matter density field overlaid with the gas velocity field. Image Credit: Illustris Collaboration.} \label{fig-illustris}
\end{figure}

 Dark matter remains invisible to us but is an essential background character. It doesn’t play much of a part in the story of the first stars aside from the origin story, though it does crop up in Section \ref{sec-global} in very much a “it was the janitor all along” kind of way. `Normal’ matter can interact with light, and so the hot gas can emit photons, removing energy from the system and therefore lowering the internal pressure of a collapsing cloud. The gravitational force can continue its work, contracting the cloud of gas further, until eventually the density of the system is so much that the hydrogen atoms fuse to form helium, producing heat and light, and stabilising the star against collapse. Our star is born, a bright ball of gas within a halo of dark matter. 

The first star has formed in our Universe, and they form all over the Universe at about the same time, because of the large-scale isotropic nature of the dark matter web. The Dark Ages draw to a close and the Cosmic Dawn begins. 

\subsection{Properties of the first stars}
Population III stars are unique in their composition, forming from primordial gas clouds free of heavy elements. This composition has important consequences for their properties.

There are three key factors that determine the properties of a star: its mass, age, and the chemical composition of the gas from which it formed, all of which are interlinked.

Small instabilities in the density field will only grow if they are above a certain critical length scale, the Jeans length ($\lambda_J$): 

\begin{equation}
    \lambda_J = c_s\sqrt{\frac{\pi}{G\rho_0}}
\end{equation}

$\lambda_J$ depends on the cosmological density ($\rho_0$), where $G$ is the gravitational constant and $c_s$ is the speed of sound in the IGM. $c_s$ accounts for the possibility that thermal pressure will counteract a collapse: 
\begin{equation}
    c_s = \sqrt{\frac{\gamma k_b T}{m}}
\end{equation}

where $\gamma$ is the adiabatic index and $k_b$ is Boltzmann's constant. When a cloud collapses, sound waves pass through the medium. If these sound waves can travel through the medium and back again before the collapse, pressure is reestablished and equilibrium maintained.

The mass at which a cloud of gas will collapse, the Jeans mass, follows as: 
\begin{equation}
    M_J = \frac{4\pi}{3}\rho_0\left(\frac{\lambda_J}{2}\right) ^3
    \label{eq-Mj}
\end{equation} 

Clouds of gas that have masses above this limit are gravitationally unstable, and able to accrete more gas and collapse under the right conditions. A cloud may be gravitationally bound, but if it cannot lose thermal heat, the internal pressure will prevent any collapse, as with dark matter halos.

By the time of first star formation, the temperature of the gas was much lower than $10^4$ K \cite{glover2005}. This limit is important, because, above this temperature, collisional excitation of electrons within atomic hydrogen, atomic helium and singly ionised helium allow the gas to cool efficiently. Already restricted to less efficient cooling routes because of the lack of metals, the low temperatures involved meant that the primordial gas could only effectively cool through the formation of molecular hydrogen, $H_2$ \cite{mcDowell1961, peebles1968} (Figure 5). \footnote{There are other cooling channels, such as lithium hydride, that are now routinely coded into simulations, however these are far less significant than $H_2$ cooling.}


\begin{figure}
\centering
\includegraphics[width=1\linewidth]{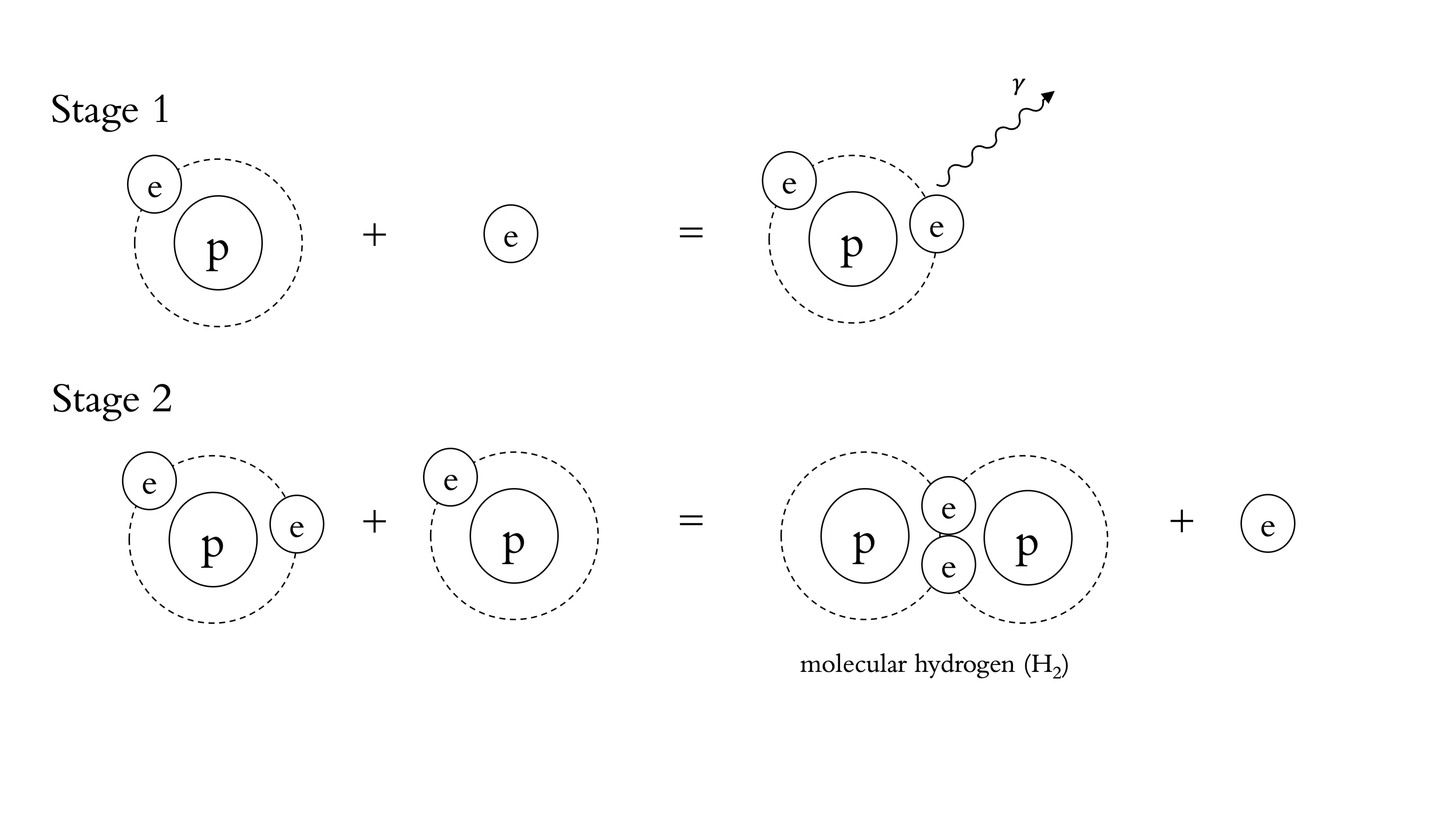}
\caption{In primordial gas clouds, the primary cooling mechanism is the formation of molecular hydrogen. Using hydrogen atoms and free electrons, this reaction releases a photon, carrying energy away from the system and allowing further collapse.}
\label{fig:H2}
\end{figure}

Under these cooling constraints, $M_J$ requires clouds of a baryonic mass $M_b > 10^5 M_\odot$ \cite{tegmark97}. This is rather large! The title of that paper is “How small were the first cosmological objects?”, but I think it should have been ‘large’ instead. With the ability to execute more complex simulations, this primordial stellar formation theory has largely held up. Many simulations demonstrated that $H_2$ cooling results in a core protostar on the order of a $M_\odot$. Over 100,000 years, though, this protostellar core accretes enough matter at a rate of about 1/10th a solar mass per year to form a massive 100 $M_\odot$ star. Initially, simulations pointed to a central, massive, lonely star, with no fragmentation in the disk \cite[e.g.][]{abel2002,bromm2004,yoshida2008}. More recently, an increased dynamic range has allowed investigation of the wider accretion disk. Multiple simulations now suggest this disk can fragment, forming binary systems \cite{turk2009, stacy2010}, and possibly even multiple star systems \cite{greif2012,greif2015}, some with orbits separated by less than 1 AU \cite{clark2011}.

How shall we search for these first stars? Do we begin by looking in our own neighbourhood? Probably not to begin with. The enormous masses of the first stars have significant implications for their lifetimes, and our chances of our observing them. The more massive a star, the more fuel to fuse, but the faster it gets through that fuel, because more massive stars require higher core temperatures and pressures to balance the increased gravitational forces. For stars the mass of our Sun, the fuel lasts about 10 billion years. For a star 100 times the mass of our Sun, a mass characteristic of Population III, the lifetime is more on the order of a few million years. Thus most of the first stars will be long dead, perhaps even all, if there is no low mass tail. Not to say noone is looking. Stellar archaeologists continue to search spectra for the absence of metal lines that would be characteristic. To survive to the present day, though, requires a stellar mass of less than 80$\%$ of the Sun. This is possible, but on the tail end of the expected mass distribution. Looking about us may therefore not be the most promising, or expedient, way of filling in our knowledge about this period on a grand scale. To really see the full range of first stars, we need to see them as they were over 13 billion years ago, before their lives so quickly ended. Luckily for us, we have a way of looking back and directly tracing the evolution of the Universe.

\section{Looking Back in Time}
	 
\subsection{Light}
Our eyes evolved to see a whole range of colours, or wavelengths, of light. That beautiful rainbow in the sky showcases about 300 nanometres of the electromagnetic spectrum. The entire electromagnetic spectrum extends from approximately 1 picometre to 1 Megametre – about 18 orders of magnitude. The visible portion is so small as to be almost insignificant, though of course it very much is not insignificant for us. Sandwiching the visible wavelengths, there are gamma-rays, and ultraviolet light at the low-wavelength end, and then microwaves and radio waves at the long wavelength end.

We are used to coming across these different kinds of light. At lunchtime, I put a ready meal in the aptly named ‘microwave’, listen to The Archers on the fittingly named ‘radio’, and perhaps flick through a comic and see Superman utilising his x-ray vision to defeat the latest threat to humanity. Different wavelengths of light serve a wide range of purposes, but they are all subject to a speed limit: $3\times10^8$ms$^{-1}$. This is so fast that our brain can’t resolve the time delay of terrestrial events. We could watch someone wave from the Moon and the signal would still only have a delay of just over 1 second. As we look farther, the time delay increases proportionally. If the Sun suddenly blinked off, we would enjoy our Summer’s day for a lovely eight minutes before it became apparent there was a problem. By observing light that has travelled longer distances, we are observing light emitted from an earlier time. Going a little further, the light we observe from Andromeda, one of our closest galaxies, is 250 million years old. Our ability to look back in time extends almost the entire way back to the Big Bang.

\subsection{Redshift}
\begin{figure}
\centering
\includegraphics[width=1\linewidth]{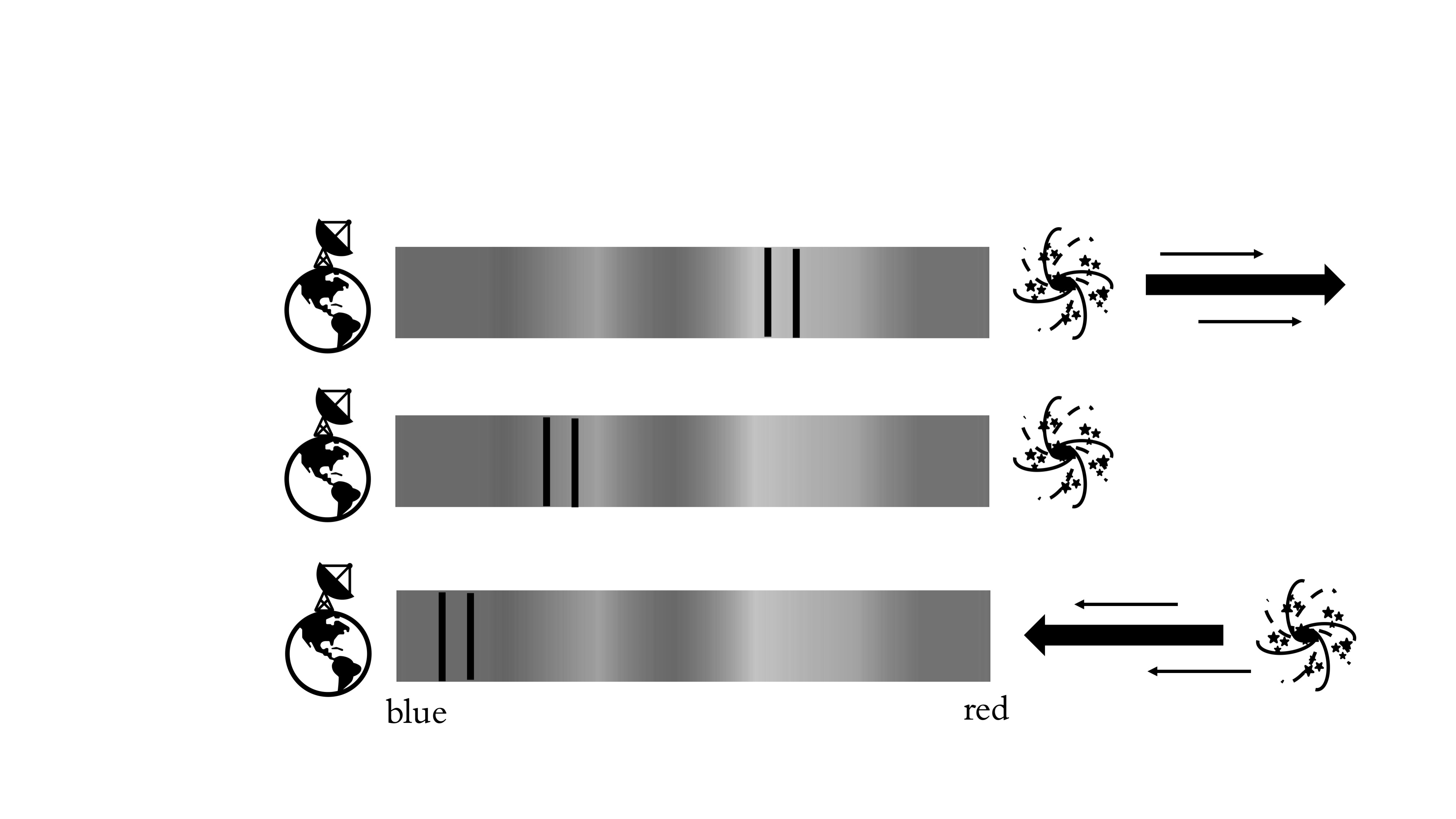}
\caption{The spectrum from a galaxy will have absorption lines relating to a variety of elements. These absorption lines will be redshifted from rest frequency if the galaxy is moving away from us, or blueshifted if moving towards us.}
\label{fig-redshift}
\end{figure}

The expansion of the Universe has been merrily progressing since the Big Bang. This has consequences for the wavelength at which we observe light, relative to the emission wavelength. The classic analogy is that of the Doppler shift we associate with the passing a fire truck roaring past you with sirens blaring. As the truck moves towards you, the sound wavefronts pile up, causing the perceived wavelength to shorten and the pitch of the siren sound you hear to increase. As the truck passes you, the siren emits the wavefronts at progressively larger spacings, causing the perceived pitch to lower. Now let’s consider galaxies moving away from us… as most of them are. As the galaxy moves away from us, flowing with the expansion of the Universe, the light emitted has steadily more separated wavefronts and we observe the light as having a longer wavelength than it was emitted at: the light is redshifted. The farther the galaxy, the faster that expansion is progressing, and so the higher the redshift of that galaxy. In this way, we can decode how far away a star or galaxy is away by the amount of redshifting of known absorption lines in its spectrum, Fig. \ref{fig-redshift}.

\subsection{Light from the First Stars}

The problem is, the first stars are so tiny in the grand scheme of things, that we simply do not have the resolving power to pick them up with any telescope, current or planned. The James Webb Space Telescope (JWST) is the flagship space telescope of humanity, and it can just about resolve fuzzy blobs of the first galaxies - there is no possibility of resolving a star \footnote{Well… there is a small possibility. In 2022, researchers found an image of an ancient star in one of the first galaxies as observed by Hubble with light emitted only 900 million years after the Big Bang \cite{welch2022}. This star had aligned perfectly with the gravitational lens created by a huge cluster of galaxies in the foreground. This was a chance alignment, and we cannot count on such an event happening again, especially for the first stars.}

We need a much stronger signal, a tracer that can tell us about the first stars. The signal we choose comes from all around the stars. We have talked about the huge abundance of hydrogen in the early Universe, some of which collapses into population III stars, and the rest of which fills the Universe. Hydrogen can emit light at many wavelengths, dependent on the atomic transition in question. There is one particular emission line that interests us: the 21 cm line. Within a hydrogen atom, the electron and proton both posses an intrinsic property called spin, which we name I and S, respectively. These spins, of value $\frac{1}{2}$, interact, leading to a splitting of the ground state \cite{field1958,field1959}:

\begin{equation}
F = S + I
\end{equation}

where $F$ can take the value 1 or 0. The $F=0$ state occurs when the spins of the proton and electron and anti-parallel, and we refer to it as a singlet as it does not split further under the influence of a magnetic field. The $F=1$ state occurs when the spins are parallel and is a slightly higher energy state. Under the influence of a magnetic state the $F=1$ state undergoes further splitting into three hyperfine state, and is called a triplet state. When an electron descends from the $F=1$ to the $F=0$ energy state, the atom releases a photon, with the properties [$\lambda$ = 21 cm, $\nu$ = 1420 MHz]. This is a `forbidden’ transition, and so an unlikely event when considering any one atom. There is a lot of hydrogen in the early Universe though. The high number density of hydrogen atoms produces a bright signal that traverses distance and time to reach us.

\section{The Dark Ages and The Epoch of Reionisation}
The Era of the First Stars can be divided into three segments: the Dark Ages, the Cosmic Dawn, and the Epoch of Reionisation (EoR) (the term `epoch of reionisation’ is almost only ever spelt with a `z’ in the literature). We can track these stages using the 21 cm signal from the hydrogen pervading the Universe, the temperature and state of which changes over time. We represent the specific intensity, $I_{\nu}$, of the 21 cm line as a brightness temperature:
 
\begin{equation}
T_b(\nu) \approx \frac{I_{\nu}c^2}{2k_B\nu^2}
\end{equation}
 
which we observe as a differential brightness temperature, $\delta_{T_b}$, relative to the CMB:
 
\begin{equation}\label{eq-Tb}
\delta T_b = 28(1+\delta)x_{\rm{HI}}\bigg(1-\frac{T_{\rm{CMB}}}{T_S}\bigg) \bigg(\frac{\Omega_bh^2}{0.0223}\bigg) \sqrt{\bigg(\frac{0.24}{\Omega_m}\bigg)\bigg(\frac{1+z}{10}\bigg)} \bigg[\frac{H(z)}{(1+z)\delta_rv_r}\bigg],
\end{equation}

where $x_{\rm{HI}}$ is the neutral hydrogen fraction, $\delta$ is the matter overdensity, and $\delta_rv_r$ is the gradient of the line-of-sight proper velocity. $H(z)$ is the Hubble `constant’, and $\Omega_m$ and $\Omega_b$ are the total matter and baryon density fractions of the Universe. The spin temperature, $T_S$, describes the population of the singlet and triplet states, which heavily depends on the coupling with the gas and CMB. If $T_S$ is equal to the temperature of the, as at extremely high redshift, then $\delta_{T_b}=0$ and no signal will be observed. As time progresses, the adiabatic cooling of the gas below the temperature of the CMB causes the brightness temperature to be seen in absorption, during the Dark Ages. By the Cosmic Dawn, the expansion of the Universe has caused the gas density to fall such that collisions are no longer effective and the 21 cm signal once again meets the temperature of the CMB and is undetectable.
 
 \begin{figure}
    \centering
    \includegraphics[width=1\linewidth]{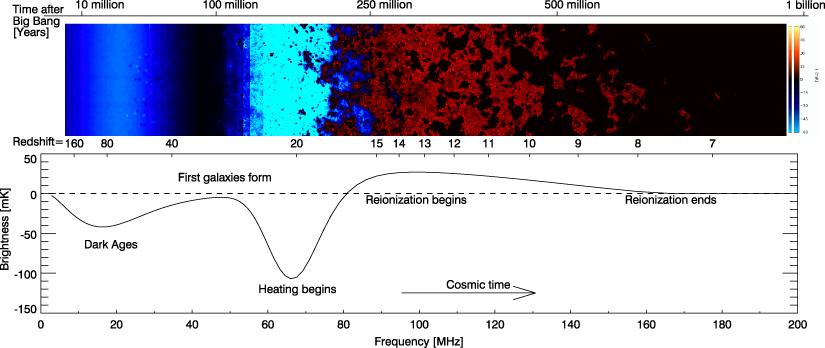}
    \caption{The brightness temperature of the cosmological 21 cm over redshift. On the top we see the tomographic signal, with the ionised bubbles forming and merging in the neutral hydrogen during the EoR. On the bottom we see the average global signal. Image Credit: Pritchard and Loeb 2012 \cite{pritchard2012}}
    \label{fig-global}
\end{figure}
 
We expect this brightness temperature to vary with time, as the gas becomes less dense with the expansion of the Universe, and the first luminous sources ionise the surrounding hydrogen Fig. \ref{fig-global}. The signal is first seen in absorption because of the fall in gas temperature below the CMB gas temperature because of the expansion in the Universe. When the first stars produce radiation capable of heating and ionizing the hydrogen gas, we mark the beginning of the EoR. Most of the gas we observe today is ionised, and so we know that within this epoch luminous sources must have carried out that process \cite{tegmark97}, earlier referenced for pioneering the calculations of primordial stellar chemistry and cloud-collapse criteria, used their calculations to work out the number of the first luminous objects that could form. Crucially, they found that a fraction of $10^{-3}$ baryons would be wrapped up as luminous objects by $z=30$, producing more than enough radiation to ionise the Universe \cite{loeb2001}. We still do not know the relative source populations though \cite{furlanetto2006,zaroubi2012,wise2019}. The number of high-redshift AGN appears too low to produce enough photons to ionise the IGM by $z\approx6$ \cite[e.g.][]{mcgreer2015} and likewise for massive population III stars if we consider realistic escape fractions.

Most studies agree that dwarf galaxies are the likely drivers of reionisation, provided a photon escape fraction exceeding 20$\%$ \cite{haehnelt2001,madau2015,fontanot2014,secunda2020}. Population III, population II, AGN and even dark matter \cite{mapelli2006} could have contributed to reionisation, driving or boosting it at different times. The question is very much open, and we need more observations within the era to constrain the many source models.

\subsection{Constraints from non-radio Observation}
The EoR may be a largely unexplored era, but we can use other observations of high-redshift Universe to provide constraints on the timing.
\subsubsection{Quasar absorption troughs}

Quasars, or `quasi-stellar objects’ are super-luminous active galactic nuclei characterised by a supermassive black hole surrounded by an accretion disc. Quasars at high redshift emit radiation that travels through an environment that will differ from the reionised environment we experience today. If the photons encounter atoms or molecules that have resonant energy gaps, absorption lines will be evident in the observed spectrum (Figure 8). The presence of hydrogen is indicated by large amounts of absorption in the quasar spectrum, because of photons exciting the Lyman-alpha transition in neutral hydrogen.

We will observe photons from distant quasars as redshifted, so photons originally emitted at high frequencies will slowly slide down the spectrum frequencies down into the Lyman-alpha frequencies. If the environment contains neutral hydrogen, then some of those photons will be absorbed, and we will measure an absorption line at the frequency of emission. This repeats with higher emission frequency photons sliding into the Lyman-alpha resonance at distances closer to us, creating another absorption line at that emission frequency. If there are more clouds of neutral hydrogen in the intervening space between us and the quasar, there will be more gaps in the spectrum - a Lyman-alpha forest. We call the region of Lyman-alpha absorption, the Gunn-Peterson trough \cite{shklovskii1964,scheuer1965,gunn1965}. The more complete the trough, the more neutral Universe the quasar photons have encountered. Using observations of when Gunn-Peterson troughs became complete, and thus when the Universe had a high neutral fraction, reionisation was thought to be largely completed by $z \approx 6$ \cite{fan2006, mcgreer2015}. Quasar observations across the sky \cite{mortlock2011, banados2018, yang2020} have led to a recent revision of this constraint. We have observed that quasars at the same redshift, but in different parts of the sky, have Gunn-Peterson troughs of differing completeness: reionisation is patchy \cite{becker2015,eilers2018, becker2018, bosman2018, bosman2021}. Simulations have suggested that this patchiness requires strong fluctuations in the ionizing UV background \cite{davies2016, aloisio2015}, other wise reionisation must end as late as $z \approx 5–5.5$ \cite{kulkarni2019}. This has been a surprise to the field, and a concerning one too, as this redshift is nearing the chosen window of EoR observation frequencies.

\begin{figure}
    \centering
    \includegraphics[width=1\linewidth]{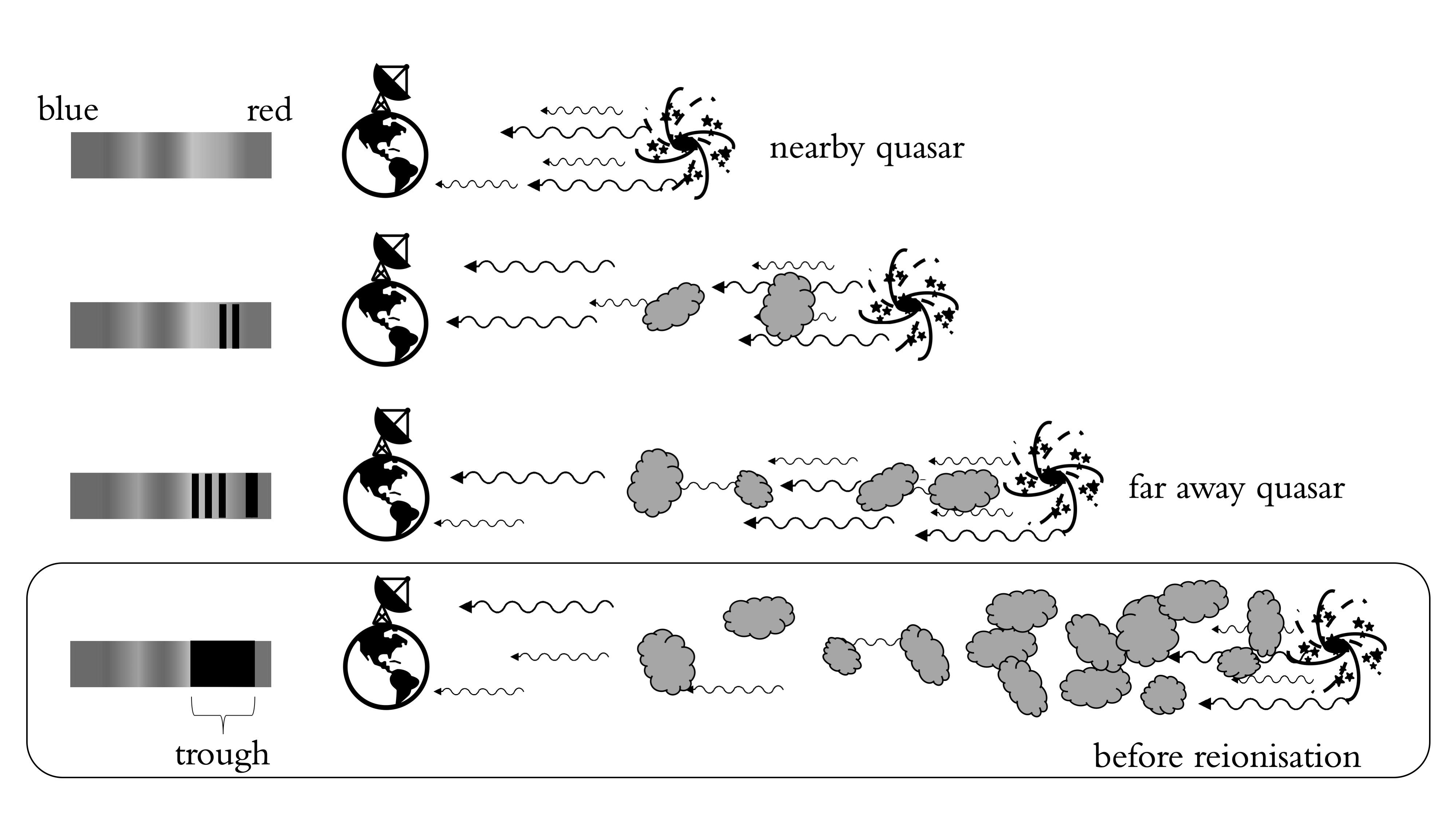}
    \caption{Quasars emit radiation across the EM spectrum, including at the Lyman-alpha transition frequency. These photons are strongly absorbed and scattered by intervening neutral hydrogen, giving us a constraint on the state of the Universe between us and the quasar.}
    \label{fig:quasars}
\end{figure}

\subsubsection{CMB}
Immediately after the Big Bang, the Universe was hot, full of interactions and obstructions for photons. It took 380,000 years of expansion for the probability of absorptions by matter particles to be low enough for the photons to travel to us largely unimpeded. The CMB is the left-over radiation from the Big Bang \cite{penzias65, dicke65}. I liken this to a jacuzzi. See a penny at the bottom and you cannot see the penny clearly at all. Switch off the jacuzzi however, remove the turmoil and turn down the energy of the system, and suddenly that light does not have such a hard time traversing the water – you’re one penny richer! The photons have immense distances and times to travel though, before our antennas pick them up. While a lot more free to move than in the very early stages of the Big Bang expansion, the photons can still be absorbed and have their path altered along cosmic time. CMB experiments like Planck and WMAP have quantified the effect of CMB photon scattering by free electrons. In a Universe that contains more free electrons, there will be more Thomson scattering. We can use this to place an integrated constraint on reionisation. If the first sources are formed and reionise the neutral hydrogen quickly, then the Universe is an ionised state for a longer period afterwards. If, however, reionisation is a more drawn-out process, then the full complement of free electrons housed within the neutral hydrogen pervading the Universe will not be released until much later. The CMB photons will not have as much opportunity to scatter across cosmic time and the Thomson optical depth ($\tau$) will be lower. Planck data suggests $\tau = 0.054 \pm 0.007$, and prefers a fast and late reionisation value with a midpoint of reionisation of $z = 7.7 \pm 0.7$ \cite{planckcosmo}. 

\subsubsection{High Redshift Galaxy Observations}
At the time of writing, JWST\footnote{https://webb.nasa.gov/} is revolutionising the breadth of astronomy, and the study of the high-redshift Universe is no exception.  On 13th July 2022, three weeks before the submission of this review, the JWST Early Release Observations were publicly released \cite{pontoppidan2022}. There followed a deluge of papers exploiting the richness of galaxies observed in the image of the SMACS J0721 lensing galaxy cluster \cite{ebeling2001}. I will not present the results in detail due to the lack of peer review, however there are now multiple robust results suggesting that massive galaxy formation started far earlier than expected. Before 13th July 2022, the farthest galaxy to be observed was at $z=10.96$, using the Hubble Space Telescope \cite{oesch2016, jiang2021}. We now have multiple candidates right up to $z=16$, at time of writing \cite[e.g.][]{labbe2022,finkelstein2022,atek2022,naidu2022,castellano2022,trussler2022,adams2022}. These surprisingly bright galaxies, observed up to only 250 million years after the Big Bang challenge our understanding of galaxy formation and evolution, and suggest that JWST will observe galaxies much farther into the Cosmic Dawn than expected.

\subsection{Radio Astronomy}

\begin{figure}
    \centering
    \includegraphics[width=0.75\linewidth]{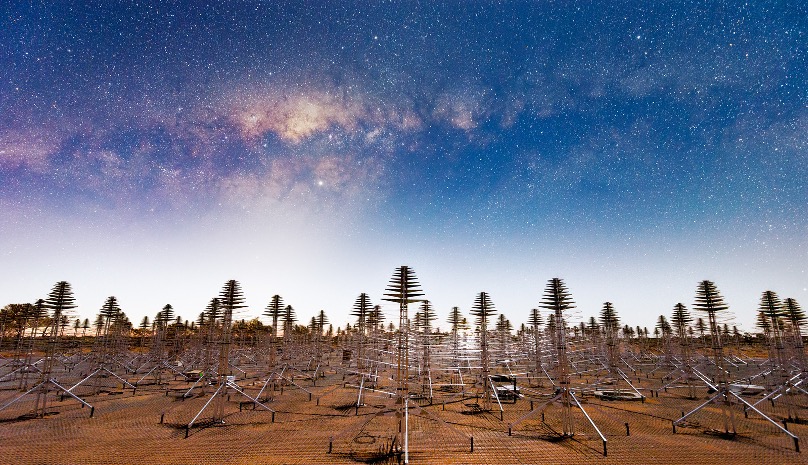}
    \includegraphics[width=0.75\linewidth]{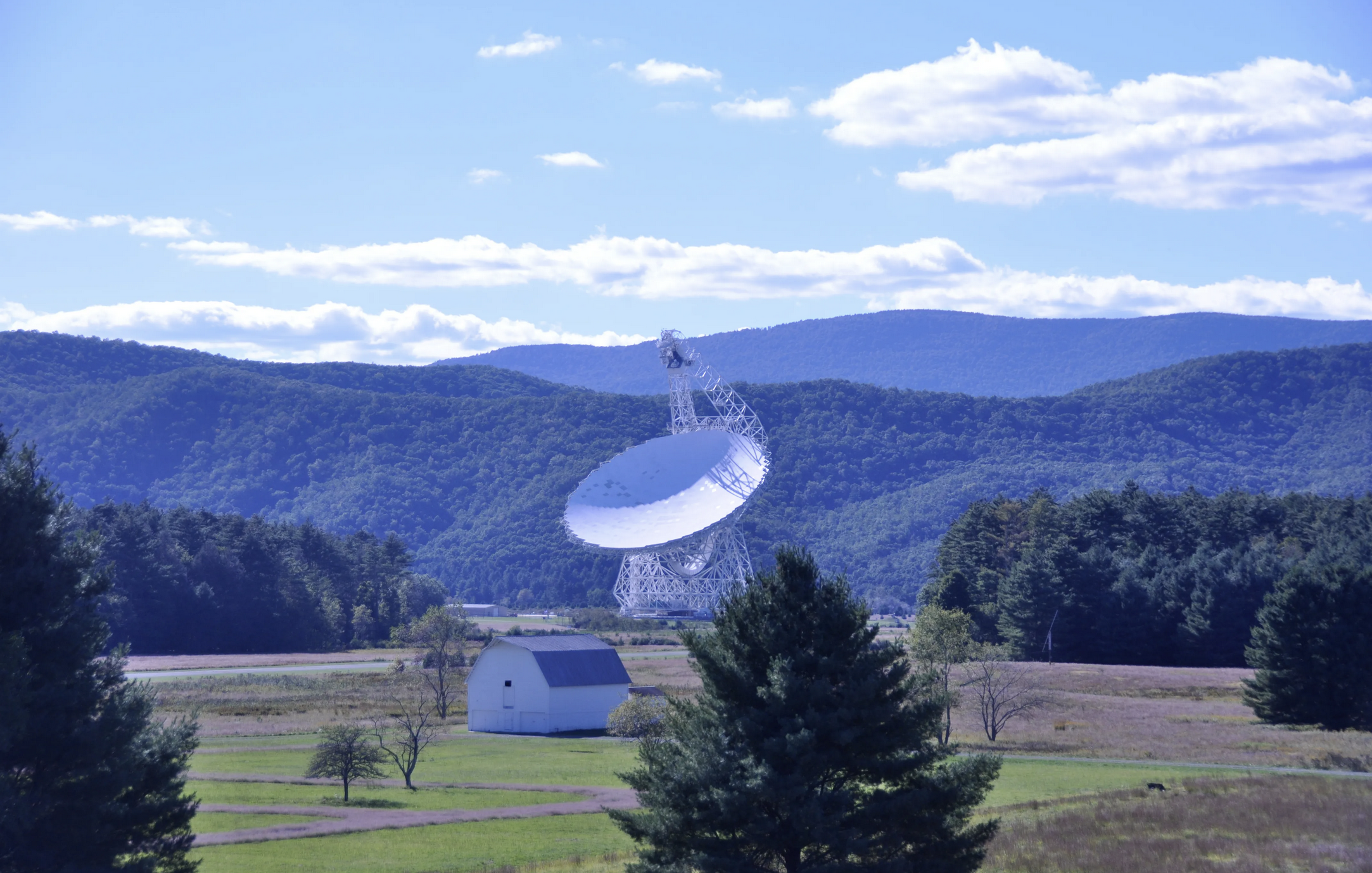}
    \includegraphics[width=0.75\linewidth]{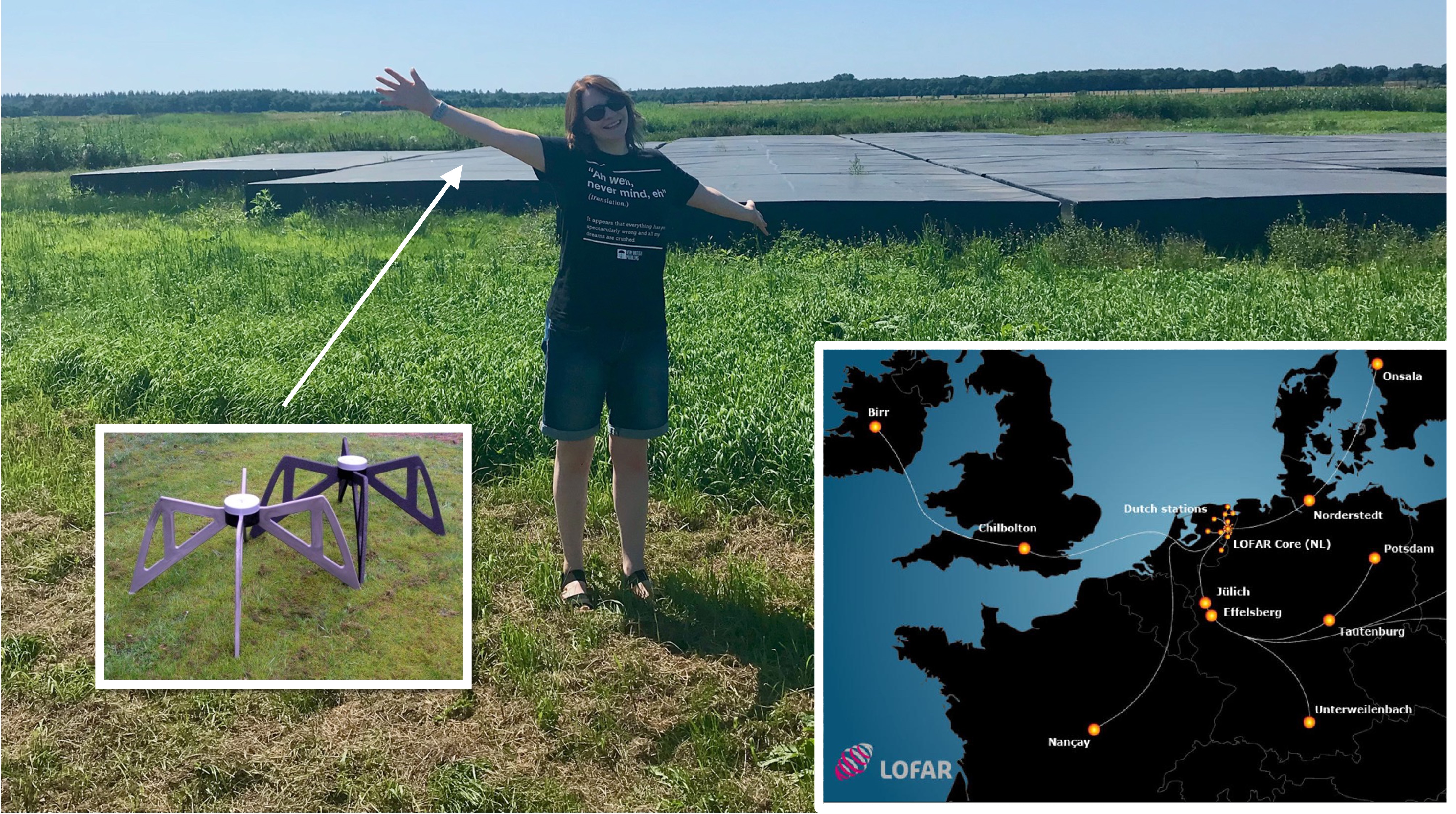}
    \caption{Top: The Milky Way overhead a test array of SKA-Low antennas. Credit: Michael Goh and ICRAR/Curtin. Middle: The Robert C. Byrd Green Bank Telescope, located at the Green Bank Observatory in West Virginia. Credit: GBO/AUI/NSF. Bottom: A station of the Low Frequency Array (LOFAR) in the Netherlands (and author), one of many stations across Europe making up the interferometer.}
    \label{fig:telescopes}
\end{figure}

None of the above constraints fill in the Era of the First Stars, or even the EoR completely. For that, we need to probe the neutral hydrogen at high-redshift using the 21 cm photon. These 21 cm photons are redshifted significantly because of their early time of emission. A 21 cm (1420 MHz) photon released at $z=8.5$, will be observed today at a frequency/wavelength of 150 MHz/2 m (During lockdown, Jodrell Bank implemented its social distancing policy using signs pointing out this wavelength). This is far out of the visible 400 nm-700 nm section of the electromagnetic spectrum by quite a long way, and sits instead within the radio spectrum, requiring a whole new kind of telescope (Figure 9).

Radio astronomy was born out of the wreckage of World War 2. Physicists and engineers built the first radio telescopes from salvaged or gifted radar (radio detection and ranging) systems that had played such a vital role in the allies’ victory. Reflecting a signal off an aeroplane and timing how long the light takes to return. This gives you an estimate of the distance of the object from the transmitting antenna. In addition, the Doppler shift introduced by the relative movement between the transmitting antenna and target can give a partial estimate of the velocity, along the line of sight joining the two parties. During the war, radar saved countless lives by providing an early warning of incoming bomber attacks, giving civilians time to take shelter. The clear transmission and detection of the signal was important, as a corrupted signal could mean the difference between noticing an incoming bomber or not. For a heart-stopping two days the English systems experienced severe radar jamming, compromising our defences. Luckily, no enemy raid ensued, which was rather odd – why jam a system if you don’t take advantage of the opportunity? Upon investigation of the timing and direction of maximum interference, the culprit was unmasked as an active sunspot crossing the solar disk, resulting in increased solar radio emission. This was not the first instance of accidental radio astronomy. The Milky Way had also been mischievous, interfering on the transatlantic telecommunications lines \cite{jansky1932,jansky1933}. Noone had the time to pursue this line of science until the war ended, but it piqued the interest of physicists such as Bernard Lovell, and amateurs such as Grote Reber \cite{reber1940b}. For the first time, we began actively listening to the Universe in radio wavelengths, and the same antennas once used for transmission, were repurposed to work only as receivers. The resolution that a telescope can achieve is proportional to its wavelength and inverse diameter. Hubble is an optical telescope, observing primarily at 0.1 - 0.8 microns. It has a mirror of 2.4m, and reaches resolutions of around 0.05 arcseconds. For an infrared telescope such as JWST, the wavelength of interest is 0.6-28 micrometres, and the mirror is 6.5m.  JWST can achieve a resolution of better than 0.1 arcseconds, a much higher resolution than Hubble can attain at longer wavelengths. For radio astronomy, the longer wavelengths in play result in significantly lower resolutions, tens to hundreds of arcseconds, and so we require much larger dishes in order to achieve high enough resolution for imaging. 

 There are some very large single radio dishes around. The steerable 76-metre dish at Jodrell Bank, UK is still observing despite being built as one of the world's first dedicated astronomical radio telescope in 1945. On the other side of the Atlantic, the 100-metre Green Bank, WV, USA stands as the world's largest fully steerable radio dish and has been scanning the skies since 2000. There is only so much weight a steerable mechanism can take, and so there have been a couple of instances of placing the dish directly in natural depressions in the ground. The most recent incarnation of this is the monumental Five-hundred-metre Aperture Spherical radio Telescope (FAST), in Southwest China. This replaces the iconic 305m diameter dish at the Arecibo observatory in the Puerto Rican jungle, which observed between 1960 and 2020. I have been there, sadly after its dramatic collapse due to faults in the cable holding the receiver above the dish. It is still a striking sight, and as I wandered underneath the dish, I enjoyed the dappled shade and jungle fauna at my feet. I was not in oppressive darkness because the dish comprises perforated aluminium sheets, letting the optical light through and providing a pleasant shade from the hot Sun. The perforations make construction and maintenance of the dish much easier, and they are far lighter and so we can construct bigger dishes. 
 
 For optical telescopes, the mirrors must be almost perfectly smooth, so why can we get away with sheets of glorified chicken wire? You can see this principle at work if you glance up from your Superman comic to see if your ready meal is done. There is a mesh on the microwave door, preventing you from having a clear view. This fine mesh has holes smaller than the wavelength of microwaves, and so the surface is effectively smooth. The microwaves reflect inwards, ensuring it is your meal that gets cooked, and not you. Optical light has a much smaller wavelength, though, and so can pass right through that mesh, allowing you to see the partially illuminated internal oven.

We can be cavalier with our definition of a radio telescope because of the wavelength of radio waves. The wavelengths we are observing allow us to have individual antennas spaced apart by metres or more, a kind of huge mesh. The more antennas in our network, which we call an interferometer, the more sensitive the telescope. The longer the longest spacing between antennas, the greater the potential resolution we can reach. That's a good thing really, since to reach the resolution of JWST, a single dish would have to extend across continents. With an interferometer, that scale is no problem. The largest-scale implantation of this so far has been the Event Horizon Telescope \footnote{https://eventhorizontelescope.org/}. Using several smaller interferometers scattered across the globe, they effectively created a telescope with a baseline the size of the Earth, and output images of the shadows of the black holes at the centre of M87 \cite{EHTM87} and our own Galaxy \cite{EHTSag}. 

There are plenty of radio interferometers aiming to make the first detection of the cosmological 21 cm signal. They span the globe: GMRT in India \footnote{http://gmrt.ncra.tifr.res.in/}, MWA in Australia \footnote{http://www.mwatelescope.org/}, HERA in South Africa (led by U.S. researchers) \footnote{http://reionization.org}, LOFAR in the Netherlands \footnote{http://www.lofar.org/}, to name the main competitors. The current generation of telescopes are known as pathfinders for a much more ambitious interferometric observatory: the Square Kilometre Array (SKA) \footnote{https://www.skao.int/}. By the late 2020s there will be 131,072 antennas in the Western Australian desert, providing an unprecedented resolution that will reach far into the Cosmic Dawn and Dark Ages. There is also another node of the telescope based in the Karoo desert of South Africa, comprising 197 dishes, observing at higher frequencies. Together these telescopes form an observatory with a wide set of science aims, including the search for traces of biological signatures in the spectra of exoplanets, the testing of general relativity using pulsar timing, and of course the investigation of the first stars and galaxies \cite[e.g.][]{braun2015}.

The interferometers utilised to make a detection of the EoR 21 cm emission do not need to cover such large baselines, but are still impressive instruments. For example, the Low Frequency Array (LOFAR) comprises 1300 antennas across Europe, and the Murchison Widefield Array (MWA) uses 4096 antennas across the Western Australian Desert. 

As with all science, discovery is an overnight success decades in the making, and these experiments have mostly been taking data for a decade or more, publishing papers about the obstacles, and the successes, from which we can all draw information for improving our own understanding. 

\subsection{The obstacles to a radio detection}
Every good story must have an obstacle to overcome, and here is ours. We have decided what we want to see, and developed ingenious new ways of enabling that, but, with victory in sight, still something stands in our way. We want to observe the first stars, tuning into light that has travelled for over 13 billion years, bearing witness to all that has changed in that time. And boy has a lot changed. When a neutral hydrogen atom emits a 21 cm photon, the Universe was still relatively simple. A sea of neutral hydrogen, with the first stars just emerging. As this 21 cm signal travels through space, the Universe evolves around it, and ever more complicated structures emerge: groups of stars, galaxies, black holes, supernovae, planets, humans with mobile phones stuck to the sides of their heads, aeroplanes, and so on. All of this emits radiation and all of it has the potential to wash out that first stars signal. We are tuning our radio antennas to be sensitive to a particular frequency: say, 150 MHz if we want to observe the 21 cm cosmological signal as emitted at $z=8.5$. The issue is, is that your antenna is not sensitive to light of cosmological origin only. The antennas will collect any radiation that arrives with a frequency of 150 MHz… and the signals are mixed and merged in a way that makes recovery of that first stars signal very tricky. Well, it is best to know thy enemy, so let’s introduce the villains of our piece.

\subsection{Astrophysical Foregrounds}
We have mapped the radiation emitted by our own and other galaxies fairly well, and the physical processes behind it are understood.

Galactic synchrotron radiation is emitted when an ultra-relativistic charged particle is accelerated in a perpendicular direction to its motion. For particles with sub-relativistic speeds, this radiation is called gyroradiation, but as soon as the speeds of the particles reach an appreciable fraction of the speed of light ($\approx85\%c$), we call the phenomena synchrotron radiation. On Earth we need expensive facilities to create synchrotron radiation, but the Universe does it for free. There are plenty of charged particles in the Galaxy, and plenty of magnetic fields to accelerate those particles. At the frequencies of the EoR experiments, the Galactic synchrotron `brightness temperature’ is thousands of Kelvin – as bright or brighter than the Sun is in the optical. 

The second largest astrophysical foreground is free-free emission. In an ionised medium, an area of the Universe where the hydrogen is ionised (an HII region), there are plenty of free ions and free electrons. These electrons and ions interacts and accelerate, releasing radiation. Before and after the interactions, however, they remain unbound (`free'). There are a lot of HII regions in the Galaxy, and so free-free emission contributes around 10$\%$ of the overall astrophysical foregrounds, a noticeably smaller contribution, but still hundreds of times larger than the cosmological signal we seek.

Our Galaxy is noisy in the radio, and so you can already see the problem with trying to detect a signal projected to be on the scale of mK in brightness. The trouble is, our villain, right in front of us all along, has rather a lot of sidekicks. The processes going on in our Galaxy are also happening in other galaxies, and the observations can give us a lot of information regarding the inner workings of the galaxy. For example, radio-loud galaxies are galaxies that produce bright jets of synchrotron radiation from their central black hole. This has become an identifying mark of a noisy AGN. This synchrotron radiation also gets in the way, making the fight one with low odds of victory… but where would be the fun be if there wasn’t a challenge?
 
Our final enemy is you. Most superhero movies I watch these days seem to have a segment where the hero struggles with their inner turmoil, trying to overcome their own past or a temptation to turn to the dark side. Well, it is here that you need to take a good look at yourself in the mirror, perhaps using the camera in your smartphone. Uh oh, you just produced an electromagnetic signal. When you turn on the TV, cook your dinner in the microwave, drive your electric car, fly in an aeroplane, use your robot lawnmower, or walk through the automatic doors at your local supermarket, just to name a few actions, you are producing radio frequency interference (RFI). Call your mum near a radio telescope and you risk wiping out the entire Galactic signal and enraging the resident astronomers. On the drive up to Jodrell Bank, a sign informs you that the Lovell telescope can pick up a phone signal on Mars. When I visited the Allen Telescope Array in California, posters reminded me that a single mobile phone can wipe out the entire Galactic signal there. 

I have to admit that when I was dealt a radio astronomy project I was a little jealous of my optical colleagues jetting off to Hawaii, Tenerife and Chile to sit on a mountaintop and stare at the sky in the dead of night. Radio astronomy is rather grittier. We don’t necessarily need high places\footnote{there is an argument that you can evade some atmospheric effects, the reason that optical needs locations at high altitude, but it is certainly not necessary to achieve the science}, but we need radio quiet locations. I have been lucky enough to visit radio telescopes across the world, and see sights you wouldn’t normally see by chance. I have tramped through long grass and across swampy ground in the depths of rural Netherlands, where the only noise came from a duck. I have driven 5 hours to the Californian alpine mountains and touched snow - in April. My latest trip took me to the Puerto Rican jungle, where I was situated in a rustic wooden cabin for a week, no Wi-Fi, no phone. RFI is a serious problem in Earthbound radio astronomy because humans love their toys. In radio astronomy, we define RFI to be interference of human-made origin. For mobile phone companies, it is the Galaxy which makes up RFI.

Luckily for us, this struggle with ourselves isn’t a major diversion, it certainly isn’t the main one. While in terms of brightness RFI can wipe out cosmological data with ease, it is also quite easy to spot and characterise. We know which frequency channels are always compromised, or we can spot them by the way they appear in the data, and we excise those channels. 

While RFI has proved surmountable, astrophysical foregrounds remain an enemy of my field. It is my job to remove them – I am a foreground mitigation scientist. The magnitude of the foregrounds is daunting, they loom over us. But they have a weakness: their spectral shape. The physical processes behind the astrophysical foregrounds are predictable and well modelled as power laws. These means that as we observe at different frequencies, the magnitude of the foregrounds change, but not the spatial form on the sky. 

In comparison, we do not see the same distribution of cosmological 21 cm. At high redshifts, the hydrogen is neutral and the 21 cm signal is the same across the sky. As the first source begin to ionise in spherical shells around them, the neutral HI takes on a bubble structure. These bubbles of ionised hydrogen grow and merge until eventually the Universe is fully ionised. Over the frequency resolution of the telescopes, the bubble distribution is uncorrelated between images. If we could see the signal clearly along one line of sight, we would see a noise-like signal. This is exploited by current foreground mitigation methods. At first, simple power laws were fit and removed in order to remove simulated foregrounds \cite{jelic2008}. However, the instruments used to observe the EoR are complex, and have varying sensitivities over frequency. This effectively changes the foreground signal over frequency, causing that characteristic coherency over frequency to weaken, and so the foreground mitigation methods relying on this characteristic to weaken too. Now, experiments use a variety of methods: blind source separation, cross-correlation, machine learning, and decomposing the signal into different scales and excising those where the foregrounds dominate, for example \cite{mertens2020,Trott2020,chapman2019}.

We detect the cosmological signal by removing the foregrounds, accounting and removing the noise. Any excess over zero should be the cosmological signal, if we have minimised all residual of the pipeline methods.

\subsection{Radio Observations}
\subsubsection{Global Experiments}
\label{sec-global}
Observations aiming to detect the cosmological 21 cm signal have been ongoing for over a decade and, as yet, we have made no definitive detection. In 2018, the media claimed a first detection of the first stars… but note I say the media. The scientists did no such thing and wrote a careful paper outlining the exhaustive process they had been through to rule out every source of instrumental and foreground residual as the cause of a strange signal that they had detected. The EDGES experiment comprised two antennas, unconnected, and one a scale version of the other, to capture different wavelengths. One of these antennas was taking data over a range of frequencies equivalent to the Dark Ages, and found a signal in absorption, at about 80 MHz \cite{bowman2018}. This is about 10 MHz later than the absorption trough in the simulation shown in Fig.\ref{fig-global}. This would have been cause enough for celebration. In addition, though, the absorption feature was about twice as deep and more flat-bottomed than any of the simulations. Clearly something was wrong with the instrument, wrong with the theories, or both. The more outlandish claim was that the data could only be explained if dark matter were interacting (which, of course, it is famous for not doing) with the gas in the early Universe and somehow making it colder than we predicted \cite[e.g.][]{barkana2018,fraser2018}. Since then, there have been many papers suggesting other ways of explaining the data. For example, we measure the 21 cm signal relative to the background radiation, which we usually assume is just the CMB. But perhaps, besides this known radiation, there is an excess radio background. This would cause the deep absorption signal not because the gas was colder, but because the background radiation was hotter \cite{ewall2018,fialkov19,mondal2020}. Or, it is quite possible that all or part of the signal is just a residual from foreground and instrument mitigation \cite{hills2018,bradley2019}. As seen in Fig. \ref{fig-global}, the global signal is an integrated value across the whole sky and is coherent with frequency, making our foreground separation techniques that rely on some element of statistical difference in coherence over frequency no longer applicable. Instead, the EDGES pipeline implemented polynomial fitting and subtraction, which we know to be less than ideal, and could have introduced a subtraction error that mimicked an absorption signal. Speculation as to whether and what EDGES detected is fast-moving, and will probably only be solved with observations from similar instruments with different environments and analysis pipelines. There are several of these in various stages of planning, commissioning and observing \cite{eastwood2019,mondal2020,acedo2022}. SARAS2 has produced initial results that contradict the EDGES results, suggesting that indeed EDGES has been the victim of some residual effect \cite{singh2022}.

\subsubsection{Interferometers} 
The global experiments mirror the COBE experiments that measured the average temperature of the CMB of the sky. Later, WMAP and Planck would measure the CMB fluctuations across the sky, allowing tighter constraints on more astrophysical and cosmological parameters. 
 
Every experiment has to have a metric for success and for our field, at least for the first detection, we have chosen the power spectrum. A power spectrum is a measure of the `power’ of a quantity across different length scales. Ideally, we would also be able to resolve the cosmological 21 cm signal, tomography. The signal-to-noise of the current generation of telescopes is likely to low to allow this.

\begin{figure}
    \centering
    \includegraphics[width=1\linewidth]{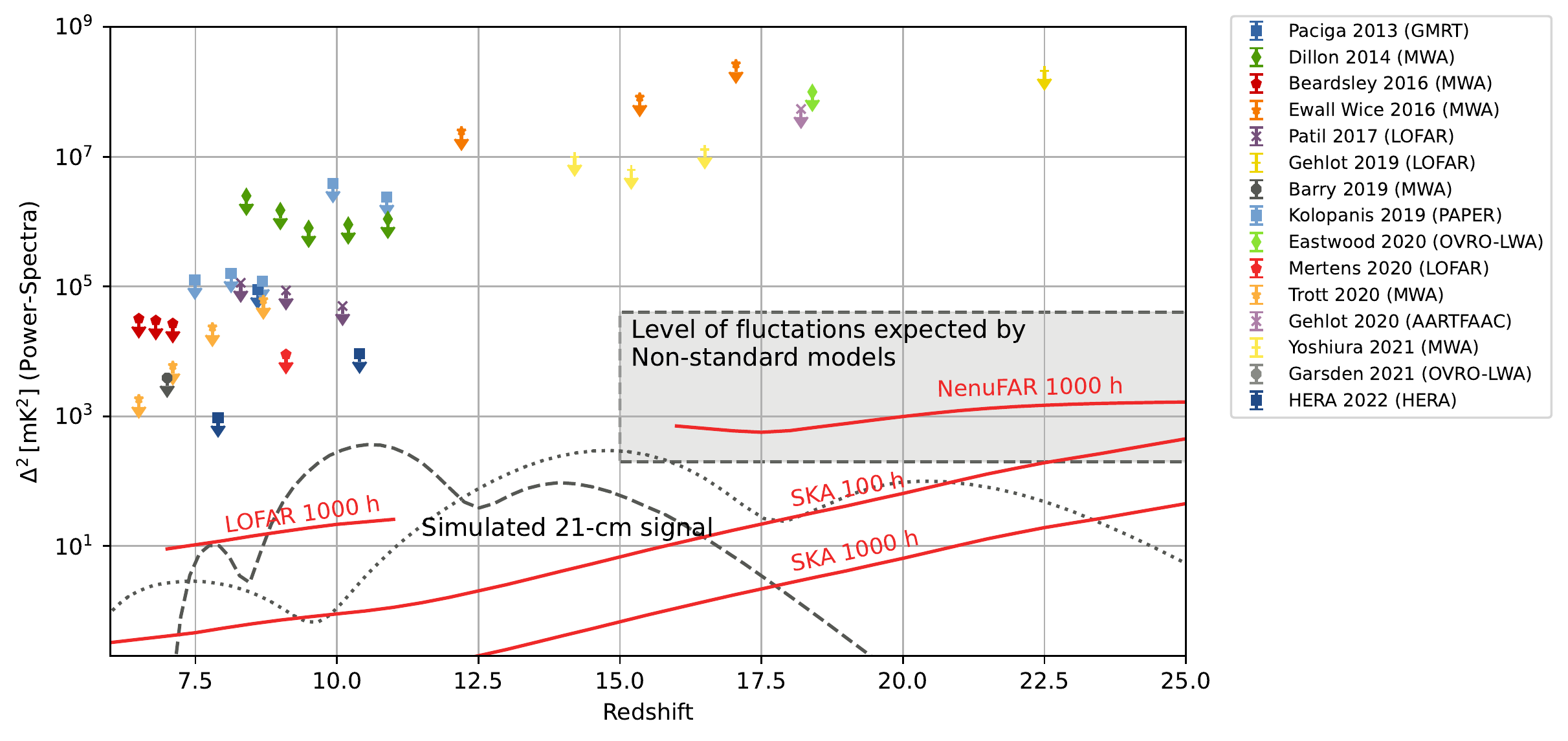}
    \caption{The upper limits of the 21 cm cosmological signal, over redshift. The experiments are all above the expected levels of a 21 cm signal, suggesting there are still significant foregrounds and other residuals to overcome. The sensitivities for several telescopes are also plotted, demonstrating that the SKA will be able to reach the expected signal-to-noise for a detection in only 100 hours of observation. Image Credit: Florent Mertens and LOFAR Collaboration.}
    \label{fig-ps}
\end{figure}

Figure \ref{fig-ps} demonstrates that the experiments are all above the expected levels of the signal to declare a detection. Currently, all experiments have only claimed upper limits, and not a detection. Starting with the observed data, the aim is to model and remove the foregrounds, and then model and remove the instrumental effects, reducing the power of the signal. Ideally, one would do such a great job of this signal cleaning, that the remaining signal would be of the order projected by cosmological signal simulations. If the cleaned signal is much higher than that expected, then it is very likely that the noise and foreground mitigation has not gone perfectly. Over time, we have refined our methods, and taken more data, in order to push our upper limits down in power, and towards the expected region of the signal. A subtlety of this scientific field is that a `detection’ will be published first as an upper limit. We will come up with an upper limit, and then over time, as a field, try and better it with new noise and foreground mitigation techniques, we will find that we cannot improve on it. Only then will we be sure enough of a detection in the face of all the obstacles.  
 
 The Dark Ages and Era of the First Stars remains largely unexplored. We have integral constraints, and a tentative detection under increasing scrutinisation, but no definitive first detection of the cosmological 21 cm signal. This has always been a waiting game, however. The complexity of our instruments, and the challenges of such a low signal-to-noise were always going to pose a challenge. Having taken the time to track down and understand the sources of interference we have encountered, we have as a field arrived at a point where the data is ready to be integrated. We have the pipelines, we have the confidence, now it is just a matter of time. And, of course, the unusual concurrence and similarity of the SKA and LOFAR/MWA generations means that lessons learned can immediately be applied to the new data analysis pipelines. There, there will be challenges and unknown unknowns too, but at least now we have a headstart.

\section{Conclusion}
We are finally uncovering the Era of the First Stars. Decades of planning and research are now all conspiring to unveil the last unknown era of our Universe: the first billion years. As the first stars burst into life, they changed the Universe, beginning the fusion of lighter elements into heavier elements, and then polluting the surrounding gas. In only one generation these first stars were mostly extinct, the gas so polluted with heavy elements that it was the familiar populations of stars around us today that formed. The detection of the 21 cm transition line from this time is one of the most promising probes. The neutral hydrogen pervading the Universe betrays what is going on under the surface, heating as the first stars form, and ionizing as the first galaxies emit enough UV photons. Following a tentative detection of this evolving signal, the field has been energised and changes on an almost monthly basis. The James Webb Space Telescope has turned out to have more of a role than initially thought. In only three weeks it has revealed galaxies from only 250 million years after the Big Bang, challenging our assumptions of massive galaxy evolution. In addition, the collection of radio telescopes around the world are all overcoming the obstacles encountered over years of observation, and should soon reach the sensitivities of an expected signal from the EoR. The unveiling of the Era of the First Stars has begun, and, over the next ten years, we can expect an exponential increase in our understanding and knowledge of this time.
\section*{Acknowledgement(s)}

EC would like to acknowledge The Royal Society for their support in providing a Dorothy Hodgkin Fellowship. In addition, The University of Nottingham for creating such a friendly and productive working environment.

\section*{Notes on contributor(s)}

Dr Emma Chapman is a Royal Society research fellow based at the University of Nottingham. Her career has involved the search for signals from the first stars, 13 billion years ago. She is a member of the LOFAR EoR team, and has lead the removal of foregrounds and noise from the data, producing the most competitive upper limits so far. Emma released her first popular science book in 2020, ‘First Light’, and has been the recipient of multiple commendations and prizes, including the Royal Society Athena Medal. Emma is a respected public commentator on astrophysical matters and gender equality issues in the sciences, regularly speaking at public events and in the media.

\bibliographystyle{tfnlm}
\bibliography{main}

\end{document}